\documentclass[showkeys,aps,twocolumn,showpacs,preprintnumbers,amsmath,amssymb,prl,letterpaper,floatfix,nofootinbib,superscriptaddress,]{revtex4-2}

\usepackage{amsmath,amssymb}
\usepackage[usenames,dvipsnames]{color}
\usepackage{graphicx}
\usepackage{subfigure}
\usepackage{soul}
\usepackage{hyperref}
\usepackage{multirow}
\hypersetup{colorlinks, linkcolor = [rgb]{0,0.0,0.75}, citecolor = [rgb]{0,0.0,0.75}, urlcolor = [rgb]{0,0.0,0.75}}

\newcommand\bef{\begin{figure}}
\newcommand\eef[1]{\label{fg:#1}\end{figure}}
\newcommand\beq{\begin{equation}}
\newcommand\eeq[1]{\label{#1}\end{equation}}
\newcommand\beqa{\begin{eqnarray}}
\newcommand\eeqa[1]{\label{#1}\end{eqnarray}}
\newcommand\bet{\begin{table}}
\newcommand\eet[1]{\label{tb:#1}\end{table}}
\newcommand\fgn[1]{Figure \ref{fg:#1}}
\newcommand\eqn[1]{Eq.\ (\ref{#1})}
\newcommand\tbn[1]{Table \ref{tb:#1}}

\begin{document}
\title{Strongly Bound Dibaryon with Maximal Beauty Flavor from Lattice QCD}

\author{Nilmani\ \surname{Mathur}}
\email{nilmani@theory.tifr.res.in}
\affiliation{Department of Theoretical Physics, Tata Institute of Fundamental
         Research,\\ Homi Bhabha Road, Mumbai 400005, India.}

\author{M.\ \surname{Padmanath}}
\email{padmanath@imsc.res.in}
\altaffiliation[Present address: ]  {The Institute of Mathematical Sciences, HBNI, Taramani, Chennai 600113, India.}
\affiliation{Helmholtz Institut Mainz, Staudingerweg 18, 55128 Mainz, Germany, \\
  {\centerline{and GSI Helmholtzzentrum f\"ur Schwerionenforschung,  GmbH, Planckstr. 1, 64291 Darmstadt, Germany.}}}

\author{Debsubhra\ \surname{Chakraborty}}
\email{debsubhra.chakraborty@tifr.res.in}
\affiliation{Department of Theoretical Physics, Tata Institute of Fundamental
         Research,\\ Homi Bhabha Road, Mumbai 400005, India.}

\preprint{TIFR/TH/22-21, MITP-22-033}

\begin{abstract}
We report the first lattice QCD study of the
heavy dibaryons in which all six quarks have the bottom (beauty)
flavor. Performing a state-of-the-art lattice QCD calculation 
we find clear evidence
for a deeply bound $\Omega_{bbb}$-$\Omega_{bbb}$ dibaryon in the
$^1S_0$ channel, as a pole singularity in the $S$-wave
$\Omega_{bbb}$-$\Omega_{bbb}$ scattering amplitude with a binding
energy $-81(_{-16}^{+14})$ MeV.  With such a deep binding, Coulomb
repulsion serves only as a perturbation on the ground state wave function of
the parameterized strong potential and may shift the strong binding
only by a few percent.
Considering the scalar channel to be the most bound for single
flavored dibaryons, we conclude this state is the heaviest possible
most deeply bound dibaryon in the visible universe.
\end{abstract}
\maketitle


Understanding baryon-baryon interactions from first principles is of
prime interest in nuclear physics, cosmology and astrophysics \cite{1967ApJ...148....3W, RevModPhys.70.303, RevModPhys.81.1773, Drischler:2019xuo}. Dibaryons
are the simplest nuclei with baryon number 2, in which such
interactions can be studied transparently. However, the only known
stable dibaryon is deuteron and the possible observation of perhaps just one more unstable 
light dibaryon [$d^*(2380)$] has recently been reported \cite{PhysRevLett.112.202301,Molina:2021bwp}
Even so,
based on the theory of strong interactions, one expects to have more
dibaryons in nature, particularly with the strange and heavy quark
contents. {\it Ab initio} theoretical investigations using lattice QCD
are well suited for studying such hadrons and indeed it can play a major
role in their future discovery.

Lattice QCD calculations of dibaryon systems are becoming more feasible now
particularly in the light and strange quark sectors \cite{NPLQCD:2010ocs,Inoue:2010es,
Luo:2011ar,Buchoff:2012ja,Berkowitz:2015eaa,HALQCD:2015qmg,Francis:2018qch,Gongyo:2017fjb,
Aoki:2020bew,Green:2021qol,Amarasinghe:2021lqa,PhysRevC.88.024003,PhysRevD.86.074514,
Berkowitz:2015eaa,Wagman:2017tmp,Horz:2020zvv}. Even so, such studies involving heavy flavors 
are limited to only a few calculations \cite{Miyamoto:2017tjs,Junnarkar:2019equ,Lyu:2021qsh,Junnarkar:2022yak}.
Among the heavy dibaryons, a system of two $\Omega_{QQQ}$ baryons ($Q \equiv c,b$)
provides a unique opportunity to investigate baryon-baryon interactions
and associated nonperturbative features of QCD in a chiral dynamics free
environment. Such a system in the strange sector have been studied using L\"uscher's finite-volume 
formalism \cite{Luscher:1990ux}, which suggested that the $\Omega$-$\Omega$ channel is weakly repulsive \cite{Buchoff:2012ja}. 
Another study using the HALQCD procedure \cite{Ishii:2006ec} suggested that the system is not 
attractive enough to form a bound state \cite{HALQCD:2015qmg}. A recent high statistics HALQCD 
study \cite{Gongyo:2017fjb} on a very large volume ($\sim$ 8 fm) claimed that such a system is 
weakly attractive and the strength of potential is enough to form a very shallow bound state. 
Although the inferences from different procedures differ, they all agree on the fact that 
the interaction between two $\Omega$ baryons is weak. Another recent HALQCD investigation 
of $\Omega_{ccc}$-$\Omega_{ccc}$ dibaryon reported a shallow bound state in the $^1S_0$ 
channel \cite{Lyu:2021qsh}. While all these investigations suggest that the interactions in two $\Omega_{QQQ}$ baryon systems are rather weak with quark masses ranging from light to charm, several lattice studies in the recent 
years on heavy dibaryons \cite{Junnarkar:2019equ,Junnarkar:2022yak} and heavy tetraquarks 
\cite{Bicudo:2012qt,Francis:2016hui,Junnarkar:2018twb,PhysRevD.100.014503} have shown that 
multihadron systems with multiple bottom quarks can have deep binding. Hence, it is very 
timely to study $\Omega_{bbb}$-$\Omega_{bbb}$ interactions using lattice QCD. Note that very little is known about it through other theoretical approaches. 
\cite{Huang:2020bmb,Liu:2021pdu,Richard:2020zxb}.

The motivation for such a study is multifold.  Theoretically it can provide an
understanding of the strong dynamics of multiple heavy quarks in a hadron. 
In cohort with results from single- \cite{Buchoff:2012ja,HALQCD:2015qmg,
PhysRevLett.120.212001,Lyu:2021qsh}, double-\cite{BEANE20111,Beane:2003da,Beane:2011iw,
Wagman:2017tmp, PhysRevC.88.024003,PhysRevD.86.074514,Berkowitz:2015eaa,
Wagman:2017tmp,Horz:2020zvv, Aoki:2020bew,Junnarkar:2019equ,Junnarkar:2022yak}, and triple-flavored dibaryons \cite{NPLQCD:2010ocs,Inoue:2010es,Luo:2011ar,
PhysRevLett.109.172001,PhysRevC.88.024003,Berkowitz:2015eaa, 
Francis:2018qch,Green:2021qol,Amarasinghe:2021lqa}, particularly those 
with heavier quarks, one would be able to build a broader picture of the baryon-baryon 
interactions at multiple scales. This can illuminate the physics of heavy 
quark dynamics in nonmesonic hadrons. A study of the quark-mass dependence of scattering parameters can further shed light 
into the dominant dynamics in different regimes. Indication of
possible promising channels on any bound heavy dibaryon from such studies can also stimulate future experimental searches for them, as in the case of heavier tetraquarks \cite{Gershon:2018gda,Ng:2021ibr,JPAC:2021rxu,Liu:2022uex}.

In this Letter, we report the first lattice QCD investigation of the
ground state of the dibaryons with the highest number of bottom (beauty)
quarks in the $^1S_0$ channel. We name it 
$\mathcal{D}_{6b} \equiv \Omega_{bbb}$-$\Omega_{bbb}$, a dibaryon formed out of
a combination of two $\Omega_{bbb}$ baryons. Using various
state-of-the-art lattice QCD utilities and methodologies, we extract
the mass of $\mathcal{D}_{6b}$ and find clear evidence for a
strongly bound state, with a binding energy of $-81(_{-16}^{+14})(14)$
MeV, and a scattering length of $0.18(^{+0.02}_{-0.02})(0.02)$ fm.  Despite
its compactness, we find the Coulomb interactions act only as a
perturbation to the strong interactions and do not change the binding
in any significant way. Upon comparison to the binding energies of other
dibaryons, {\it e.g.} 2.2 MeV of deuteron, and other strange or heavy
dibaryons \cite{Lyu:2021qsh,Junnarkar:2019equ}, we conclude 
$\mathcal{D}_{6b}$ to be the most deeply bound heaviest possible dibaryon in our visible universe.



The lattice setup that we use here is similar to the one used in 
Refs. \cite{Mathur:2018epb,Junnarkar:2018twb} and we discuss it below.

\bef[t]
\centering
\vspace*{-0.07in}
\includegraphics[scale=0.35]{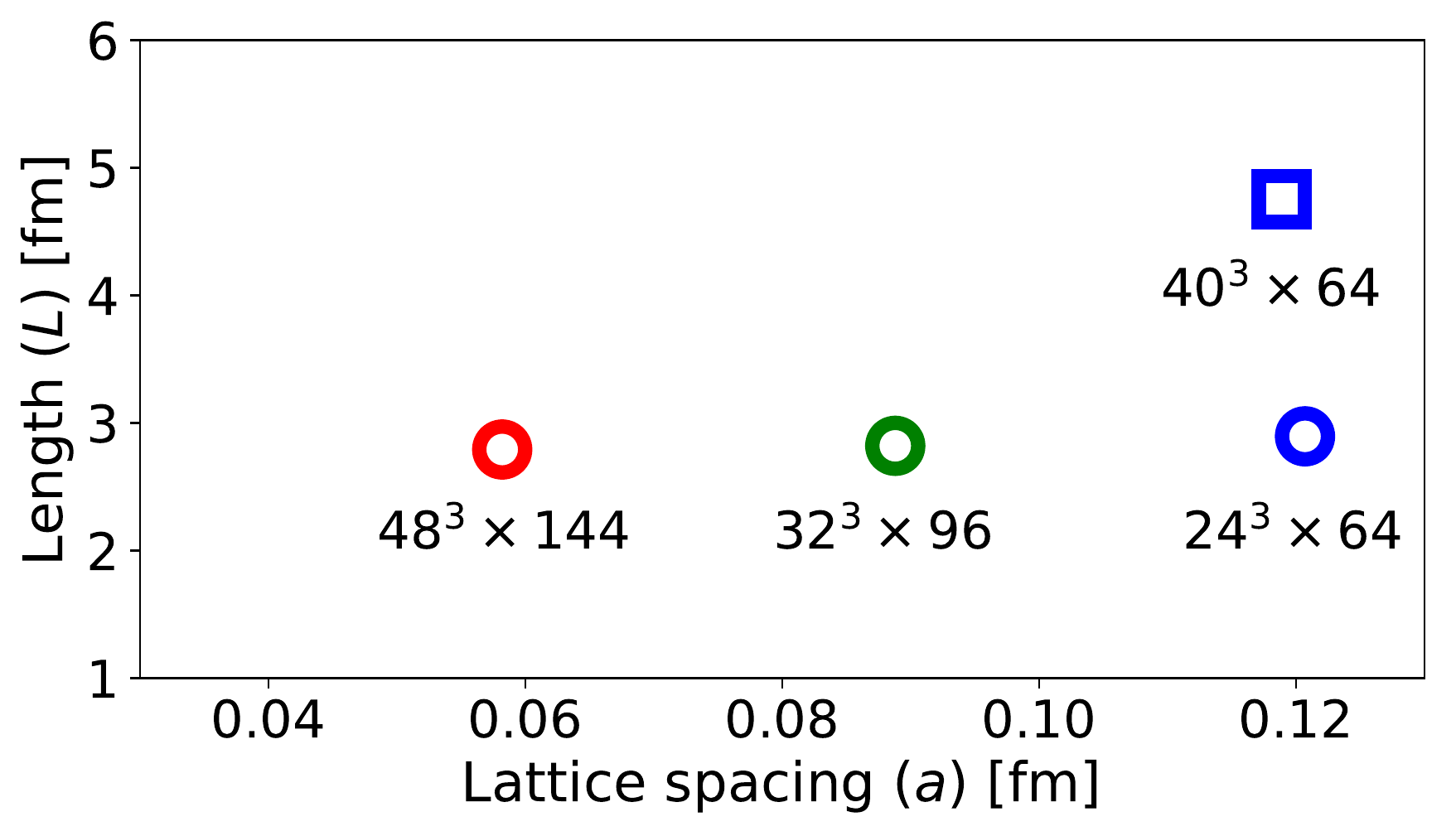}
\vspace*{-0.05in}
\caption{\label{fig:pars}Lattice QCD ensembles, with sizes $N_s^3\times N_t$, used in this work. Here $L = N_sa$ is the spatial extent of the lattice.
}
\eef{}
\noindent{\it{Lattice ensembles:$-$}}
We employ four lattice QCD ensembles with dynamical $u/d,~s$ and $c$ quark fields, generated by the MILC Collaboration~\cite{Bazavov:2012xda} with highly improved staggered quark (HISQ)
fermion action~\cite{Follana:2006rc}, as shown in Fig. \ref{fig:pars}. Lattice spacings are determined using $r_1$ parameter \cite{Bazavov:2012xda}, which are found to be consistent with the scales obtained through Wilson flow~\cite{Bazavov:2015yea}.
%

\noindent{\it{Bottom quarks on lattice:$-$}} 
Since the bottom quark is very heavy, we use a nonrelativistic QCD
(NRQCD) Hamiltonian \cite{Lepage:1992tx}, including improvement
coefficients up to $\mathcal{O}(\alpha_sv^4)$
\cite{Dowdall:2011wh}. Quark propagators are calculated from the
evolution of NRQCD Hamiltonian with Coulomb gauge fixed wall sources
at multiple source time-slices.  We tune the bottom quark mass using
the Fermilab prescription for heavy quarks~\cite{ElKhadra:1996mp} in
which we equate the lattice-extracted spin-averaged kinetic mass of
the $1S$ bottomonia states with its physical value
\cite{Tanabashi:2018}. Such a tuning was also used in
Refs. \cite{Mathur:2018epb,Junnarkar:2018twb,Junnarkar:2019equ} and
was found to reproduce the physical value of the hyperfine splitting
of $1S$ bottomonia.


\noindent{\it{(Di)baryon interpolators:$-$}}
For the single $\Omega_{bbb}$ baryon, we use the quasilocal nonrelativistic 
operator with $J^P=3/2^+$, as was used in Ref. \cite{Buchoff:2012ja}. This operator was 
constructed by the LHPC Collaboration and is listed in Table VII of Ref. \cite{Basak:2005ir} and 
also detailed in Ref. \cite{Suppl}. For extracting the ground state mass we assume 
only $S$-wave interactions in two baryon systems where the overall state is 
antisymmetric under the exchange of two baryons. Denoting components of the $J$=3/2 
$\Omega_{bbb}$ operator ($\mathcal{O}_{\Omega_{bbb}}$) with $\chi_m$, $m$ being
the azimuthal component of $J$, we construct the $\Omega_{bbb}$-$\Omega_{bbb}$ dibaryon operators as,
\beq
\mathcal{O}_{\mathcal{D}_{6b}}(x,t) = \chi_m(x,t)~[CG]^{mn}~\chi_n(x,t).
\eeq{Eq:Domega1}
Here, $[CG]^{mn}$ are the relevant spin-projection matrix constructed out of 
the appropriate Clebsch-Gordon coefficients. The $J=0$ dibaryon operator that
 we employ in this work, is given by \cite{Buchoff:2012ja, Suppl},
\beq
\mathcal{O}_{\mathcal{D}_{6b}^{J=0}} = \frac12~\Bigl[\chi_{\frac32}\chi_{-\frac{3}2}+\chi_{-\frac12}\chi_{\frac12}-\chi_{\frac12}\chi_{-\frac12}-\chi_{-\frac32}\chi_{\frac{3}2}\Bigr].
\eeq{Eq:Domega12}



Using these baryon and dibaryon operators ($\mathcal{O}_{\Omega_{bbb}}$ 
and $\mathcal{O}_{\mathcal{D}_{6b}}$) we compute two-point correlation 
functions between the source ($t_i$) and sink ($t_f$) time slices, 
\beq
C_{\mathcal{O}}(t_f-t_i) = \sum_{\vec{x}_f} e^{-i\vec{p}.\vec{x}_f}\langle 0 | \mathcal{O}(\vec{x}_f,t_f)\bar{\mathcal{O}}(t_i)|0 \rangle.
\eeq{Eq:Domega2}
At the sink time slice, we use several different quark field smearing 
procedures to identify the reliable ground state plateau and quantify possible 
excited state contamination (see Ref. \cite{Suppl} for more details).
Ground state masses for the single and the dibaryon are obtained by 
fitting the respective average correlation function with a single 
exponential at large times ($\tau = t_f-t_i$).
\bef[b]
\centering
\includegraphics[scale=0.35]{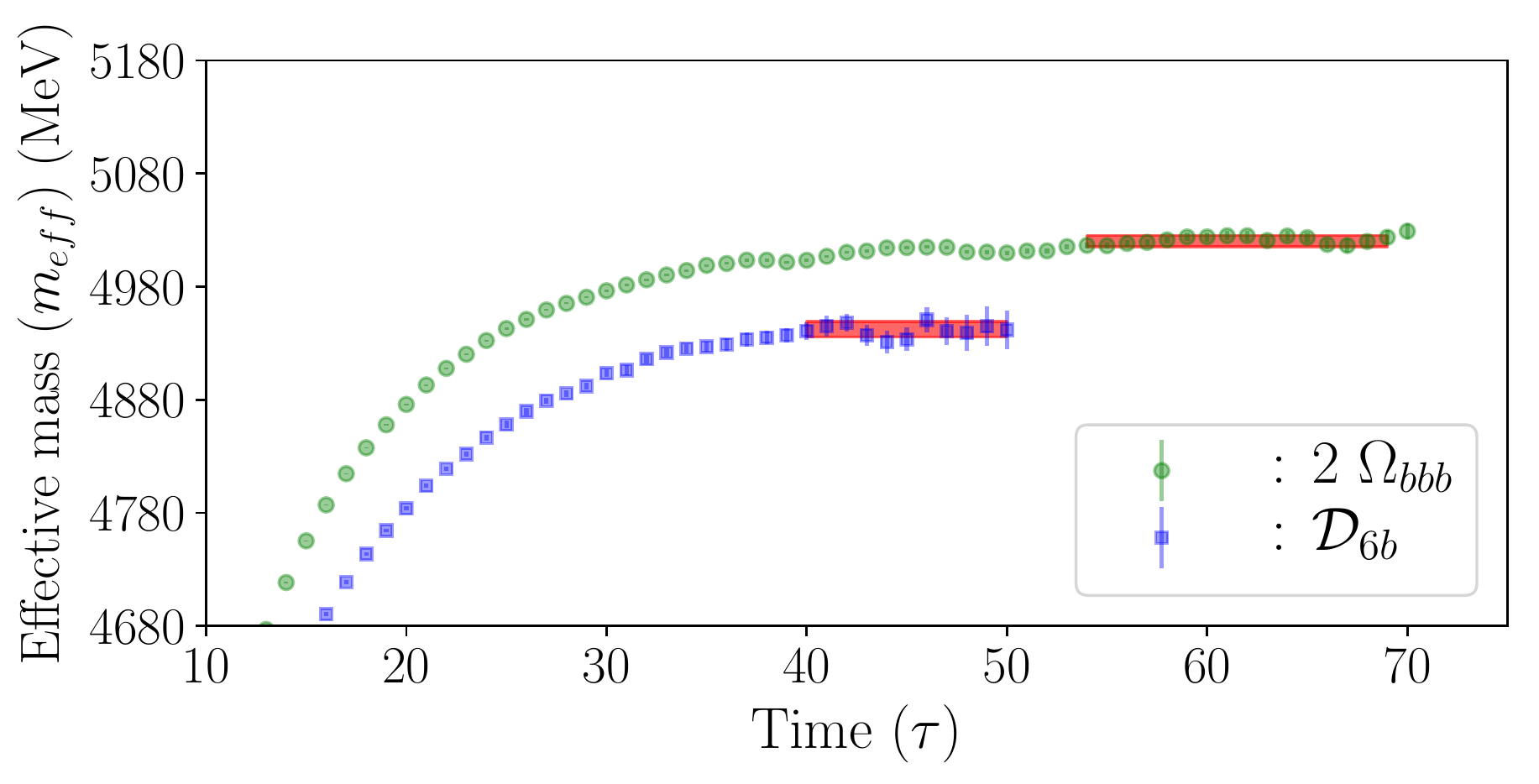}
\vspace*{-0.09in}
\caption{\label{fig:eff-mass}
  Effective masses corresponding to the ground states of the noninteracting two-baryon and dibaryon correlators on the finest lattice ensemble determined from wall-to-point correlation functions. An energy gap between them is clearly visible at all time slices. The solid bands show the fit estimates and fit windows.
}
\eef{fg:effmass}

While determining mass in a lattice calculation it is often useful to plot the effective mass, 
defined as $m_{{eff}}a = log[\langle C(\tau) \rangle/\langle C(\tau+1) \rangle]$, to show 
the signal saturation and justify the time window to be chosen in the exponential fit. In 
Fig. ~\ref{fig:eff-mass}, we present the effective masses for $C^{2}_{\Omega_{bbb}}$ (green 
circles) and $C_{\mathcal{D}_{6b}}$ (blue squares) on the finest ensemble ($a \sim 0.06$ fm) using wall quark sources and point quark sinks. 
We make the following observations from this result: (i) The signal in the effective masses 
saturates well before the noise takes over, and hence one can reliably extract the respective 
ground state masses. (ii) The signal in the noninteracting 2$\Omega_{bbb}$ level survives 
until large times. This is because 2$\Omega_{bbb}$ level is obtained from the single baryon 
$\Omega_{bbb}$ correlator that decays with an exponent of $M_{\Omega_{bbb}} < M_{\mathcal{D}_{6b}}$, 
and hence can propagate further than the $\mathcal{D}_{6b}$ state. (iii) Most importantly, 
it is quite evident that there is a clear energy gap between the ground state energy levels 
of the noninteracting two-baryon and the dibaryon systems at all times. This clearly shows 
that the ground state mass of dibaryon $M_{\mathcal{D}_{6b}}$ is smaller than that of the 
non-interacting level $2M_{\Omega_{bbb}}$. We find similar energy differences for all the 
ensembles and we discuss the results below. Based on the $t_{\mathrm{min}}$
dependence of the fits, which are discussed in Ref. \cite{Suppl}, we make our final choices 
for the fit ranges and uncertainties arising out of such choices.

In order to gauge the extent of excited state contaminations in our estimates, 
we carry out two additional calculations: one with a wall source and a Gaussian-smeared 
sink \cite{Davies:1994mp,Wurtz:2015mqa}, and the other with a wall source and 
spherical-box sink \cite{Hudspith:2020tdf}. The results are detailed in the 
 Supplemental Material \cite{Suppl}. We find that results are clearly consistent between different 
measurement setups and validate our estimates. We pass the results from all 
these different smearing procedures through 
the scattering analysis, as discussed below, to determine uncertainties related to the excited state contamination. 
Moreover, an effective mass analysis using Prony's method 
\cite{Prony1795,Fleming:2004hs,Kunis2015AMG} and a lattice setup with 
displaced baryons \cite{Suppl}, further reinforce the findings of two clearly separated energy levels as in Fig. \ref{fig:eff-mass}  \cite{Suppl}. 


Next we calculate the energy difference between the ground state of the dibaryon 
($\mathcal{D}_{6b}$) and the noninteracting two baryons ($2~{\Omega_{bbb}}$)
\beq
\Delta E = M_{\mathcal{D}_{6b}} - 2 M_{\Omega_{bbb}}.
\eeq{eq:dE}
In Table~\ref{tb:dE}, we present $\Delta E$ for all the lattice ensembles. We quote 
the average of various fitting and smearing procedures as the central value of energy 
splittings in separate ensembles. The largest deviation in these energy splittings extracted from different procedures is taken as the systematics 
related to the excited state effects. We find $\Delta E$ to be always {\it negative} and several 
standard deviations ($\sigma$) away from zero. This observation on multiple ensembles, 
with three different lattice spacings,  two different volumes, and different 
energy extraction procedures lead us to unambiguously conclude that there is an energy 
level below the threshold.
\bet[h]
\centering
\begin{tabular}{@{\hspace{0.3cm}}c @{\hspace{0.3cm}} @{\hspace{0.3cm}} c @{\hspace{0.7cm}}| @{\hspace{0.7cm}}c @{\hspace{0.3cm}}  @{\hspace{0.3cm}}c @{\hspace{0.3cm}}}\hline \hline 
  Ensemble & $\Delta E$ & Ensemble & $\Delta E$\\
  \hline \hline
  $24^3 \times 64$ & $ -61(11)$ & $40^3 \times 64$ & $ -62(7)$ \\
  $32^3 \times 96$ & $ -68(9)$  & $48^3 \times 144$ & $ -71(7)$\\
  \hline 
\end{tabular}
\caption{\label{tb:dE}Energy difference $\Delta E = M_{\mathcal{D}_{6b}} - 2M_{\Omega_{bbb}}$ 
in MeV on different ensembles.}
\eet{tb:delE}



\noindent {\it Scattering analysis:$-$} To establish the existence of a state from these energy levels 
in terms of pole singularities in the $\Omega_{bbb}\Omega_{bbb}$ $S$-wave scattering amplitudes 
across the complex Mandelstam $s$-plane, we use the generalized form of finite-volume formalism 
proposed by M. L\"uscher \cite{Luscher:1990ux}. For the scattering of two spin-3/2 particles in 
the $S$-wave leading to a total angular momentum and parity $J^P=0^+$, the phase shifts $\delta_0(k)$ 
are related to the finite-volume energy spectrum via L\"uscher's relation: 
\beq
k~cot[\delta_0(k)] = \frac{2Z_{00}[1;(\frac{kL}{2\pi})^2)]}{L\sqrt{\pi}}.
\eeq{luscher} 
Here, $k$ is the momentum of $\Omega_{bbb}$ in the center of momentum frame and 
is given by
\beq
k^2 = \frac{\Delta E}{4} (\Delta E+4M_{\Omega_{bbb}}^{\mathrm{phys}}), 
\eeq{Ecalc}
where $\Delta E$ is the energy differences listed in \tbn{tb:delE}, and $M_{\Omega_{bbb}}^{\mathrm{phys}}$ is the mass of $\Omega_{bbb}$ in the continuum limit. The $S$-wave scattering amplitude is given by $t = ({\mathrm{cot}}\delta_0 - i)^{-1}$, and a pole in $t$ related to a bound state happens when $k~{\mathrm{cot}}\delta_0 = -\sqrt{-k^2}$. 
We parameterize $k~{\mathrm{cot}}\delta_0 = -1/a_0$, where $a_0$ is the scattering 
length. The scattering analysis is performed following the procedure outlined in Appendix B of Ref. 
\cite{Padmanath:2022cvl}, such that the best fit parameters are constrained to satisfy Eq. (\ref{luscher}). 
To estimate the systematic uncertainties from the lattice cut-off effects, we perform several different fits involving different subsets of the four levels with $k~cot\delta_0$ parameterized either as a constant or as a constant plus a linear term in the lattice spacing. All of the fits indicate the existence of a deeply bound state. We find that the best fit corresponds to the one that considers all energy levels and incorporates the lattice spacing $a$ dependence of the scattering length with a linear parameterization $k~{\mathrm{cot}}\delta_0 = -1/a_0^{[0]} - a/a_0^{[1]}$. We present this as our main result, 
leading to a $\chi^2/d.o.f= 0.7/2$, with the following best fit parameters and binding energy 
\beqa
a_0^{[0]} = 0.18(^{+0.02}_{-0.02})~\text{fm}, &\quad&  a_0^{[1]} =-0.18(^{+0.18}_{-0.11})~\text{fm}^2,\label{scatlen} \\
\text{and~~} \Delta E_{\mathcal{D}_{6b}} &=& -81(^{+14}_{-16})~\text{MeV}.
\eeqa{beDO}
In \fgn{pcot1}, we present details of our main results. On the top pane, the analytically reconstructed finite-volume energy levels (black stars) from best fit parameters in Eq. (\ref{scatlen}) can be seen to 
be in agreement with the simulated energy levels (large symbols), indicating quality of fit. In the middle pane, we plot $k~{\mathrm{cot}}\delta_0$ versus $k^2$ in units of the energy of the threshold. The orange dashed curve is the bound state constraint $\sqrt{-k^2}$ and the red solid line is the fitted $k~{\mathrm{cot}}\delta_0$ in 
the continuum limit. The crossing between these two curves, highlighted by the magenta symbol, is the bound state pole position in $t$.
In the bottom pane, we present the continuum extrapolation of binding energy leading to the value in \eqn{beDO} compared with the simulated energy levels at the respective lattice spacings. The magenta symbol represents the binding energy in the continuum limit, with thick error representing the statistical and fit window error. The thin error includes the systematics related to excited state effects added in quadrature.
\bef[h]
\centering
\includegraphics[height=7.5cm,width=8.8cm]{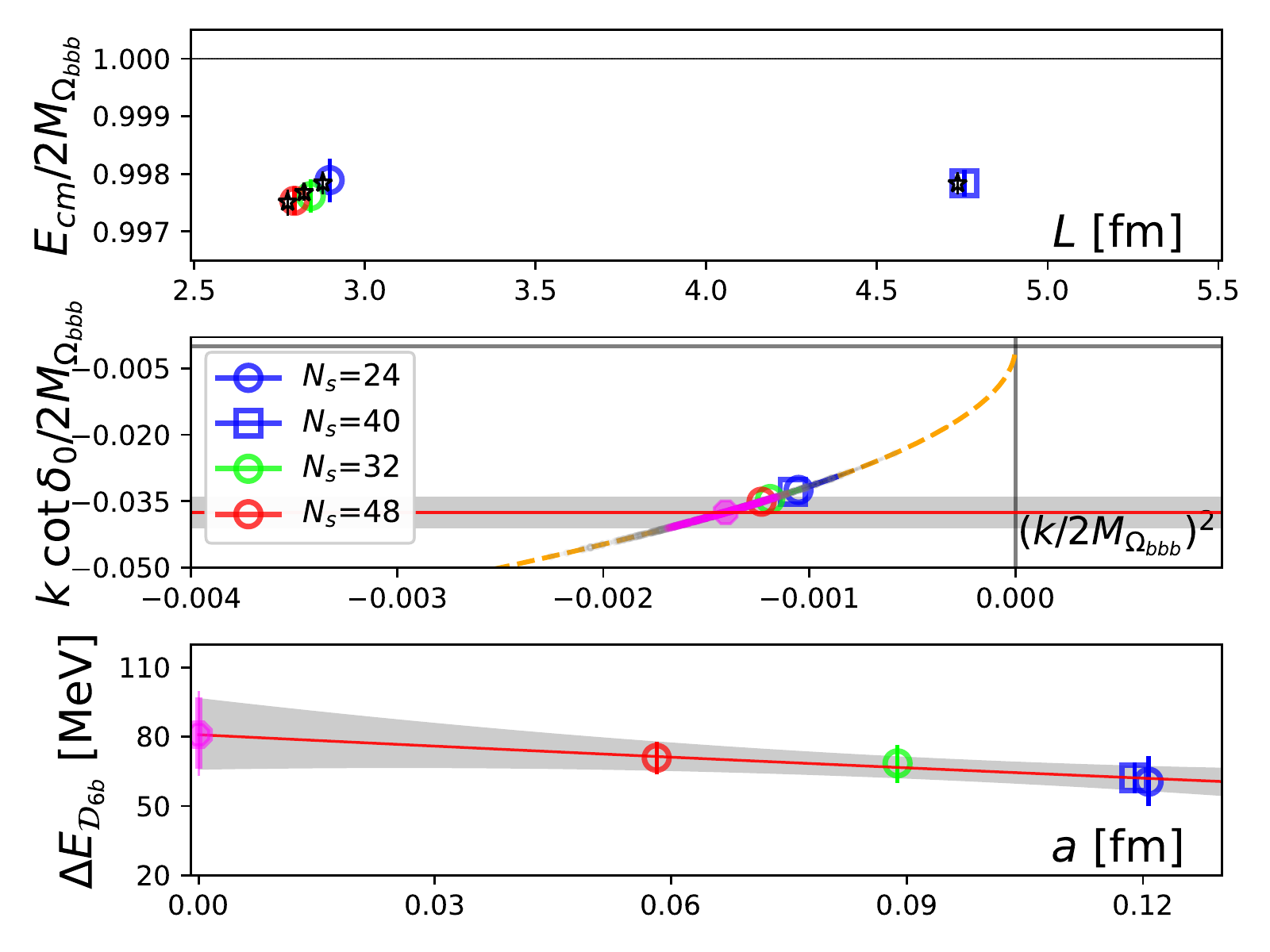}
\caption{Results from the finite-volume scattering analysis. Top: Comparison of the simulated energy levels (large symbols) with the energy levels (black stars)
analytically reconstructed using Eq. (\ref{scatlen}), indicating the quality of the scattering analysis fit. 
Middle: $k~{\mathrm{cot}}\delta_0$ versus $k^2$ in units of energy of the threshold ($2M_{\Omega_{bbb}}$) and information on poles in $t$ indicated by magenta symbols. Bottom: Continuum extrapolation of the binding 
energy in \eqn{beDO} determined from fitted scattering amplitude in Eq. (\ref{scatlen}).}
\eef{pcot1}



\noindent {\it Coulomb repulsion:$-$} With two units of electric charge in the system, the effect of 
Coulomb repulsion on the binding energy of this dibaryon could be important. To gauge that, we 
perform an analysis, as in Ref. \cite{Lyu:2021qsh}, and detail that in the Supplemental Material \cite{Suppl}. We model the strong interactions 
between two interacting $\Omega_{bbb}^{-}$ baryons with a quantum mechanical multi-Gaussian attractive potential, constrained  to match the binding energy $-81(_{-16}^{+14})$ MeV that we find in this work. In Fig. \ref{fig:coul1}, we present the model potentials for strong and Coulombic interactions and also their 
combination, together with the radial probabilities of the ground state wave functions in the strong and combined 
potentials. Evidently, the Coulombic potential hardly affects the strong interaction potential in the length scales
where the ground state probabilities peak and infer that it serves only as a perturbation. The associated maximum change in binding energy is found to be between 5 and 10 MeV. 

\bef[t]
\centering
\includegraphics[scale=0.35]{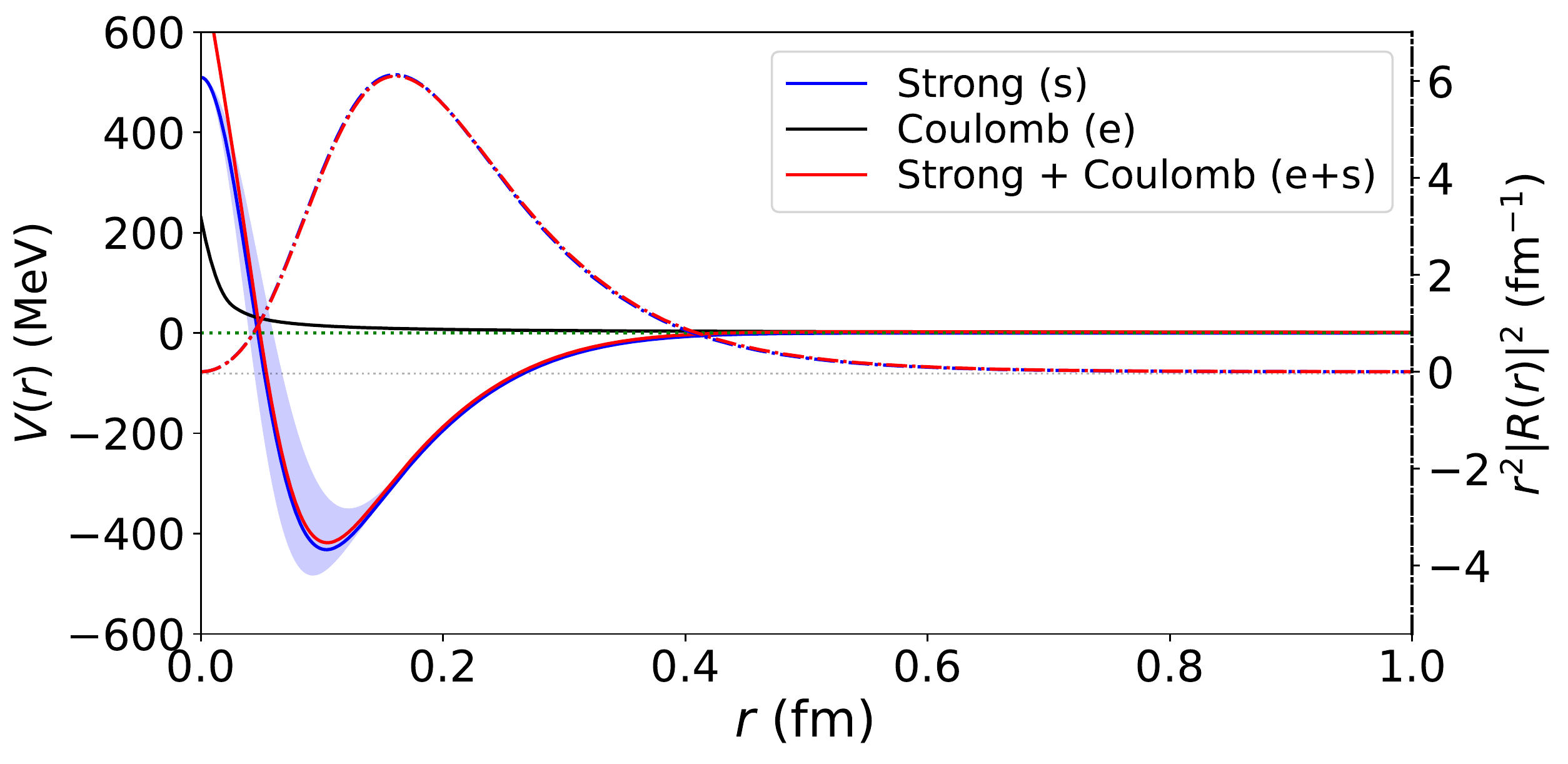}
\caption{\label{fig:coul1} Coulomb ($V_e$), the parameterized strong potentials ($V_s$) and their sum are shown by the black, blue and red curves, respectively. The shaded region represents the variation of $V_s$ with respect to its parameters. $V_e$ is evaluated at a {\it rms} charge radius equal to the {\it rms} radius of the $V_s$ ground state. The radial probability densities of the ground state wave-functions of the strong and combined potentials are shown by the dashed-dotted curves.}
\eef{}



After addressing the systematic errors along with excited state contaminations \cite{Suppl} the final value of the dibaryon mass is determined by adding 
$\Delta E_{\mathcal{D}_{6b}}$ [$-81(^{+14}_{-16})(14)$ MeV] with the two-baryon mass $2 M_{\Omega_{bbb}}$. Since the $\Omega_{bbb}$ baryon mass is unknown  we use its lattice 
extracted value. To this end, we perform continuum extrapolation of the energy splitting $M_{\Omega_{bbb}} (a)- {3\over 2} M_{\overline{1S}}(a)$, and then add  
3/2$M_{\overline{1S}}^{\mathrm{phys}}$, with $M_{\overline{1S}}^{\mathrm{phys}} = 9445$ MeV \cite{Zyla:2020zbs}, to that. Thus we arrive at $M_{\Omega_{bbb}}^{\mathrm{phys}}= 14366(7)(9)$ MeV, which is consistent with other lattice results \cite{Brown:2014ena}. Using that, we obtain $M^{\mathrm{phys}}_{\mathcal{D}_{6b}} = 2M_{\Omega_{bbb}}^{\mathrm{phys}} + \Delta {E}_{\mathcal{D}_{6b}} = 28651(^{+16}_{-17})(15)$ MeV. Possible effects of Coulomb repulsion are included in the systematic errors.



\noindent{\it Error budget:$-$} Finally we address the possible sources of errors in 
this calculation. We use a lattice setup with 2+1+1 flavored HISQ fermions where 
the gauge fields are Symanzik-improved at $\mathcal{O}(\alpha_sa^2)$, and the NRQCD 
Hamiltonian has improvement coefficients up to $\mathcal{O}(\alpha_sv^4)$. Such 
a lattice setup has shown to reproduce energy splittings in bottomonia with 
a uncertainty of about 6 MeV \cite{Suppl}. Note that here we are 
calculating the energy difference in which some of the systematics get reduced. 
For the dibaryon ground state in the finite volume, 
statistical, excited-state-contamination, and fit-window errors are the main 
sources of error. The energy levels are extracted using single exponential fits to the correlation 
functions from rigorously identified ground state plateau regions \cite{Suppl}.  Correlated averages of various fitting intervals are considered to arrive at conservative 
fitting-window errors. Statistical and fit window errors are added in quadrature, and 
then convolved through the L\"uscher's analysis and continuum extrapolation. The excited state contamination 
is determined from differences in the continuum limit estimates from the scattering 
analysis using results from different sink smearing procedures followed.
However, it still would be worthwhile to investigate excited state uncertainties more precisely in future variational calculations. Other possible 
sources of errors are related to the continuum extrapolation fit forms, scale setting, 
quark mass tuning and electromagnetic corrections that together are found to be 12 MeV
in such energy splittings, as detailed in the Supplemental Material \cite{Suppl}. Various errors are finally added in quadrature, yielding 
a total error of about 20\% for the binding energy. Our results and inferences are robust up to the statistical and systematic uncertainties 
we have determined.

\noindent{\it Summary and Outlook:$-$} In this Letter, using lattice QCD we present a first investigation of the dibaryons in which all six quarks have bottom flavor  and find a deeply bound dibaryon 
($\mathcal{D}_{6b} \equiv \Omega_{bbb}$-$\Omega_{bbb}$) in the $^1S_0$ channel. Following L\"uscher's formalism, we determine the relevant scattering amplitude, and  after considering possible systematic uncertainties \cite{Suppl}, we identify a bound state pole with a binding energy $-81(^{+14}_{-16})(14)$~MeV relative to the threshold $2M_{\Omega_{bbb}}$. The mass of {$\mathcal{D}_{6b}$} dibaryon corresponding to this pole is found to be 28651($^{+16}_{-17}$)(15)~MeV. Although this dibaryon is expected to be compact, we find the Coulomb repulsion within this dibaryon acts only as a perturbation to the strong interactions and may shift the mass only by a few percent. The use of complementary measurements and analysis procedures in identifying the real ground state plateau ensure the robustness of our results. Our results provide intriguing evidence for the existence of the bound $\mathcal{D}_{6b}$ state, and it would surely motivate both phenomenological studies of its detection as well as follow-up lattice QCD studies investigating hard-to-quantify excited-state uncertainties more precisely.

It is interesting to observe that the interactions between similar baryons using different procedures 
at the strange and charm quark masses are found to be very weak \cite{Buchoff:2012ja,HALQCD:2015qmg,Gongyo:2017fjb,Lyu:2021qsh}. 
Note that a clear consensus on such systems with possible near threshold features requires complementary 
investigations of the same system with same high statistics ensembles but with different procedures. In comparison 
with the light and strange sectors, the binding energy of multiquark hadrons involving more than one bottom quark are predicted to be large \cite{Bicudo:2012qt,Francis:2016hui,Junnarkar:2018twb,
  PhysRevD.100.014503,Karliner:2017qjm,Eichten:2017ffp,Junnarkar:2019equ}. In this work we also observe the similar pattern in the $\Omega_{bbb}\Omega_{bbb}$ channel.
Taken together a common interesting pattern is emerging that the presence of more than one bottom quark enhances the binding in multihadron systems, which needs to be understood thoroughly including the quark mass dependence of scattering parameters.

Although a direct identification of $\mathcal{D}_{6b}$ dibaryon is a long way to go, our results 
on this heavy dibaryon, particularly because of its deep binding, will provide a major impetus in experimental searches for heavy quark exotics. 
Very much like the discovery of $\Xi_{cc}$ leading to predictions of various possible heavy multiquark 
systems \cite{Karliner:2017qjm}, the discovery of doubly bottom baryons would be an important step in filling up 
the blanks higher up in the hadronic reaction cascade bringing prospects for discovering various bottom quark exotics,
including $\mathcal{D}_{6b}$. Given the recent excitements in the search for new heavy exotics \cite{LHCb:2021vvq,LHCb:2019kea,LHCb:2020bwg} with multiple theoretical proposals and ideas \cite{Gershon:2018gda,Ng:2021ibr,JPAC:2021rxu,Liu:2022uex}, 
it is highly anticipated that substantial efforts, both on the theoretical as well as experimental fronts, would be steered and accelerated in this direction in the coming years.


\begin{acknowledgments}
This work is supported by the Department of Atomic Energy, Government of India, under Project Identification Number RTI 4002. We are thankful to the MILC collaboration and in particular to S. Gottlieb for providing us 
with the HISQ lattice ensembles. We thank D. Mohler for a careful reading of the manuscript. We thank  the authors of Ref. \cite{Morningstar:2017spu} for making the {\it TwoHadronsInBox} package utilized in this work and in particular Colin Morningstar for his help with the package. We also thank T.~Doi, L.~Liu, T.~Luu, M.~Pappagallo, and S.~Paul for discussions. Computations are carried out on the Cray-XC30 of ILGTI, TIFR. N. M. would also like to thank A. Salve, K. Ghadiali, and P. M. Kulkarni for computational supports. 
\end{acknowledgments}


\bibliography{OmegabDb}
\section*{Supplementary information on calculation details and error analysis -- \\
{\it Strongly Bound Dibaryon with Maximal Beauty Flavor from Lattice QCD}}
\section*{Calculation details}

\noindent{\bf{Calculating the energy differences}:} \,\,  
The Euclidean two-point correlation functions in Eq. (3) at large source-sink separations $\tau$
are fitted with single exponentials of the form
\begin{equation}
\label{Eq:eff_split}
\langle C (\tau) \rangle  = W_{0}e^{-E_0\tau},
\end{equation}
using correlated $\chi^2$ and maximum likelihood estimators to extract $E_0$ and $W_0$.
In Figure \ref{fig:maxl} we present such a result showing the projections of posterior probability distributions 
of the parameters $E_0$ and $W$ demonstrating the reliability of the fits for the example of $\mathcal{D}_{6b}$ correlation
functions in the finest ensemble.

\bef[h]
\centering
\includegraphics[height=7.5cm,width=8cm]{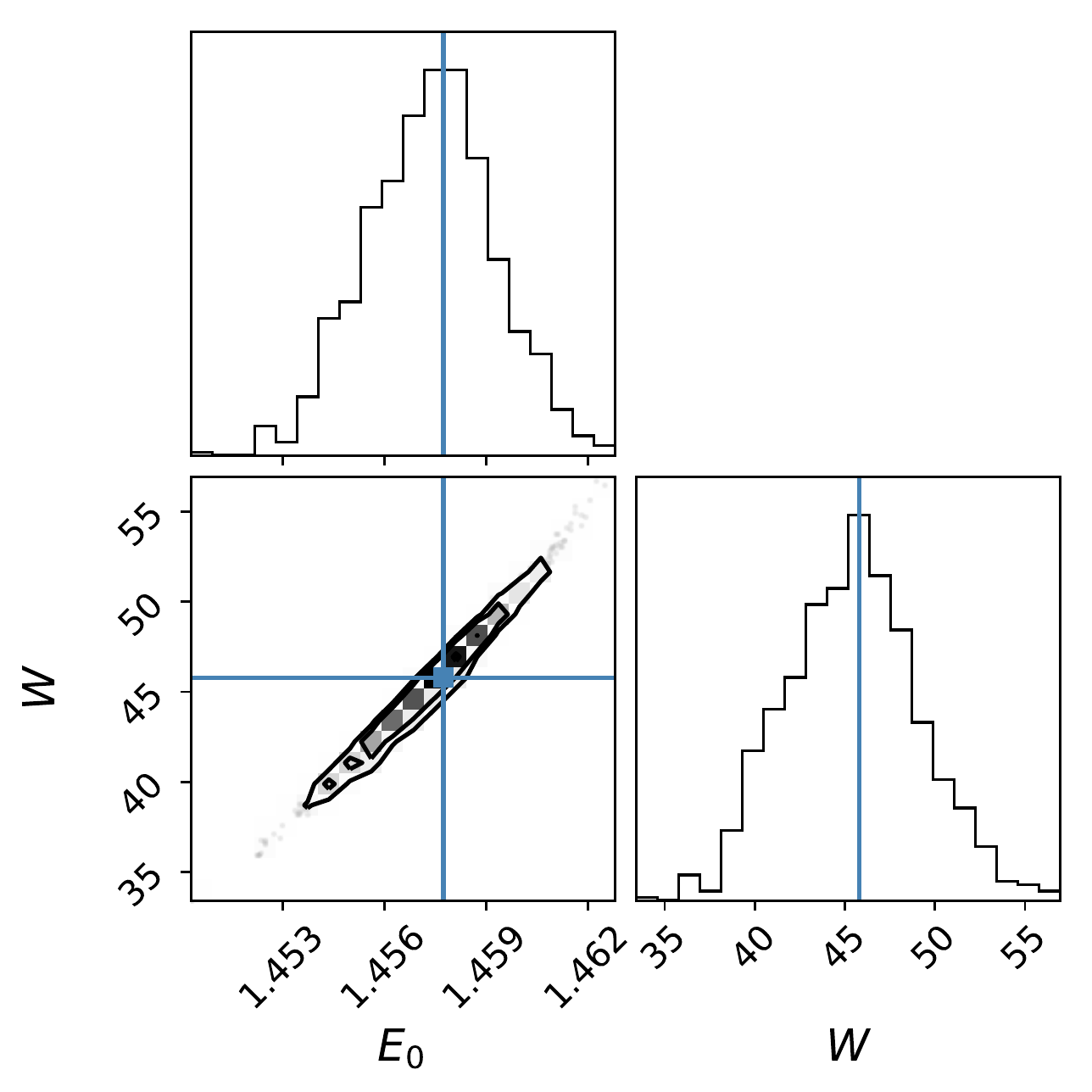}
\caption{\label{fig:maxl}  
The multivariate distribution of fitted parameters $E_0$ and $W$ (bottom left) and their respective univariate distributions
(Top left) and (Bottom right).}
\eef{fg:maxl}

In order to quantify the uncertainties arising from the choice of fitting window ($\tau_{min}, \tau_{max}$), 
we do the following. First choose a $\tau_{max}$ as large as possible with a good signal-to-noise ratio. Then the $\tau_{min}$ is varied over a range to determine the stability of $E_0$ estimate and 
a $\tau_{min}$ value is chosen where a clear plateau is observed. A conservative estimate taking account of an 
uncertainty on this choice is arrived at using a correlated average over neighboring $\tau_{min}$ values in the plateau.
In  Figure \ref{fig:tmin-plot-all}
we present the $\tau_{min}$ dependence for all the fits along with the 1-$\sigma$ statistical errors for the chosen
fit window (blue bands), and the final estimate considering the uncertainty from the chosen fitting window (magenta bands). In both figures, we present the 
estimates for single baryons on the left and for dibaryons $\mathcal{D}_{6b}$ on the right. 
These estimates are then utilized to arrive at the energy differences in Eq. 4  and Table I of the main text.

\bef[h]
\centering
\hspace*{-0.06in}\includegraphics[scale=0.53]{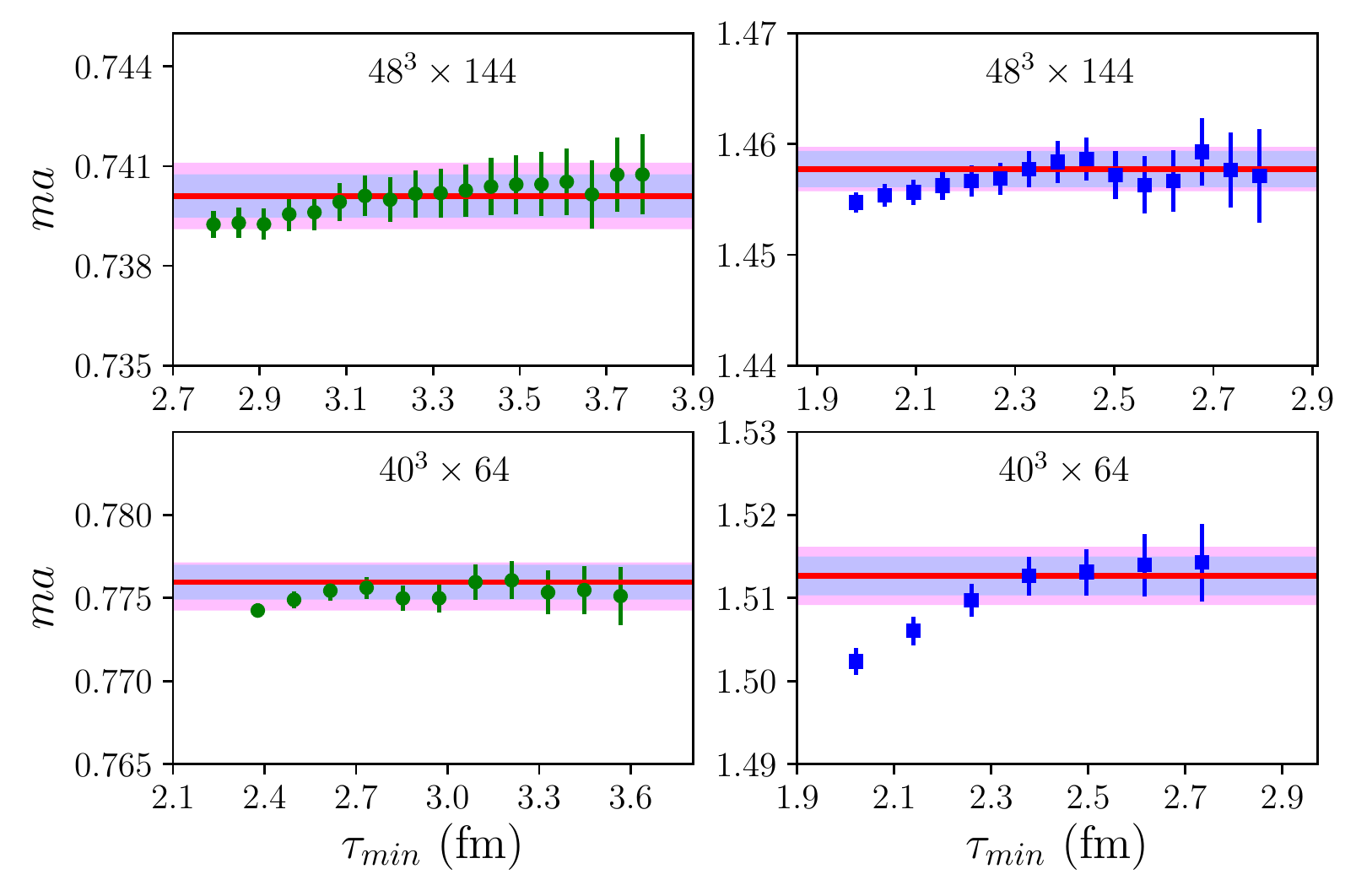}\\
\hspace*{-0.06in}\includegraphics[scale=0.53]{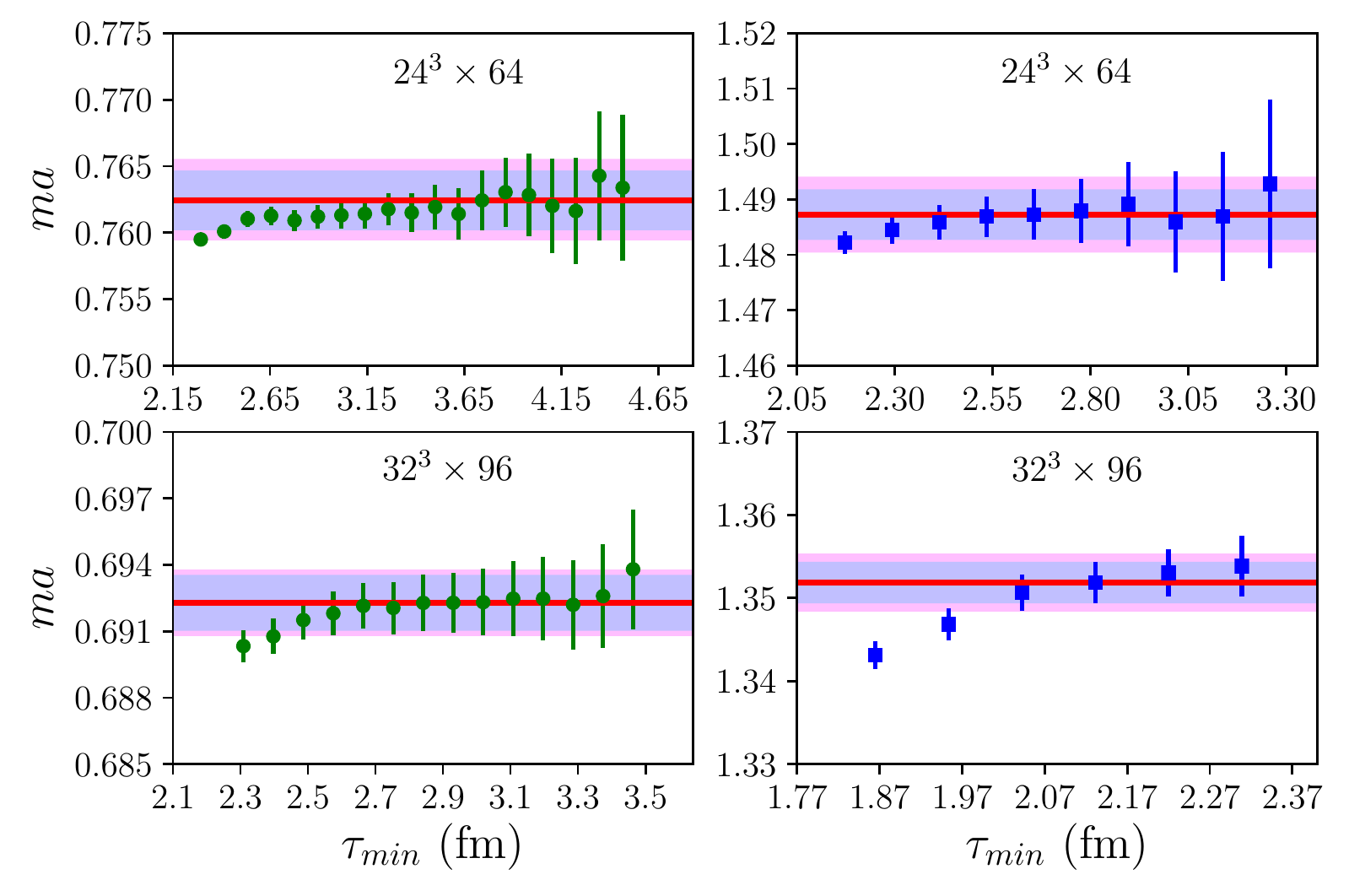} 
\caption{\label{fig:tmin-plot-all} Fit results for the ground state masses for different fit-windows corresponding to various choices of minimum time ($\tau_{min}$).}
\eef{}



\section*{Error Analysis}
The main source of error in a lattice QCD calculation for a multi-hadron system  arises from 
the rapid decrease of the signal-to-noise ratio in the correlation functions \cite{BEANE20111}. 
In heavy hadrons, this is somewhat mitigated due to the presence of heavy quarks. In this 
calculation since all the valence quarks are of bottom flavor and no chiral dynamics is involved,
it is expected to have a relatively better signal-to-noise ratio than that of other dibaryons. 
Nevertheless, different systematics need to be addressed, particularly those arising from
the contamination of excited states and from the lattice discretization,  
to arrive at a reliable estimate for the binding energy of the $\mathcal{D}_{6b}$ dibaryon. 
We discuss various relevant systematics involved in our calculation below. 

\vspace*{0.1in}

\noindent{1. {\it{Extraction of the ground state masses}}:}

\noindent Coulomb gauge fixed wall sources are utilized for the quark fields, which has been widely used 
over many years for calculations involving NRQCD and is known to produce good ground state plateau at a large source-sink separation.
We also average the correlation functions over multiple 
source time-slices to improve the statistical uncertainties. In addition to this, we follow the 
procedure outlined in the previous section to include fit-window uncertainties, and arrive at 
the final energies and energy differences presented in the main text.

\bet[h]
\begin{tabular}{@{\hspace{0.3cm}}c | @{\hspace{0.3cm}}c |@{\hspace{0.3cm}}c }\hline \hline
  $\mathcal{O}_{\Omega_{bbb}}$ & $|s_1s_2s_3\rangle$ & $|J, J_z\rangle$ \\\hline
  $\chi_1$ & $|111\rangle$   & $|3/2,+3/2\rangle$ \\
  $\chi_2$ & $|112\rangle_S$ & $|3/2,+1/2\rangle$ \\
  $\chi_3$ & $|122\rangle_S$ & $|3/2,-1/2\rangle$ \\
  $\chi_4$ & $|222\rangle$   & $|3/2,-3/2\rangle$ \\
  \hline
\end{tabular}
\caption{Four rows of the spin 3/2 baryon operator in the $^1H^+$ irrep. $\chi_i$ refers 
to different components of the $\mathcal{O}_{\Omega_{bbb}}$ operator. $s_1$, $s_2$, and $s_3$
refer to the spin components of the quark constituents. With NRQCD action, it can either be 
up or down, referred by 1 and 2 respectively in the second column. The subscript $S$ refer
to the symmetrized form of the operator. The third column shows the total spin and the azimuthal 
component of each row of the operator.} 
\eet{Obbb_op}
For $J=3/2$ $\Omega_{bbb}$ baryons, we utilize the most symmetric operator ($^1H^+$) with 
rows given in \tbn{Obbb_op}, also expressed in Eq. (7) of Ref. \cite{Buchoff:2012ja}. 
It was observed from studies which had previously utilized this operator for the studies of 
$\Delta$ \cite{Edwards:2011jj,Edwards:2012fx}, $\Omega$ \cite{Edwards:2012fx,Buchoff:2012ja}, 
$\Omega_{ccc}$ \cite{Padmanath:2013zfa} and $\Omega_{bbb}$ \cite{Meinel:2012qz} baryons, that 
the ground state has the largest overlap with this operator, and the ground states for these 
singly-flavored baryons are best determined with this operator. It is also observed that the 
radial excitations for the decuplet baryons are very high in energy arising from higher partial
waves. For $\Omega_{bbb}$, the first radial excitation is observed to be $>400$ MeV above the 
ground state $\Omega_{bbb}$ from the lattice calculation in Ref. \cite{Meinel:2012qz}. Any 
reminiscent effects from those should be reflected as significantly different approach to 
the energy plateau in the effective energy plots and the $t_{min}$ dependence plots in different 
lattice QCD ensembles. We observe in all our ensembles, which vary in lattice spacing and 
the volume, the signal plateauing commences from approximately the same physical temporal extent ($\gtrsim$ 2.5 fm ).

For the dibaryon, we utilize the $S$-wave projected two-baryon interpolating operator
expressed in Eq. (2) in the main text, as was also utilized in Ref. \cite{Buchoff:2012ja}.
In our case, the action being non-relativistic we are limited to the two-baryon operator built 
purely out of $^1H^+$ baryon operators, which is also observed to be the best choice for the ground 
state determination. We also note that the large time approach to the energy plateau in this case 
also occurs around the same physical temporal extent ($\gtrsim$ 2 fm ) across all our ensembles, 
which confides the reliability of our energy estimates. Nevertheless, to ensure it further, we 
perform a set of additional calculations, both for single and dibaryon systems, in various setups 
to arrive at a conservative estimate on possible excited state contaminations in our results. We 
describe that below.

\vspace*{0.1in}

\noindent{2. {\it{Excited state contamination}}:}

\noindent The use of wall smearing at the quark source and no smearing at the quark sink is 
an asymmetric setup in building two point correlation functions. This results in an
unconventional rising-from-below behavior of the effective energies, as a result of 
competing overlap factors with different signatures for different states having same 
quantum numbers. Consequently, there could be low lying plateaus at early times that 
mimics the real ground state plateau. To remove this complication, we perform a set of 
calculations with different source-sink setup, study the asymptotic behavior in search of 
a universal estimate for the ground state energy level along with a reliable estimate for 
possible uncertainty from excited state contaminations. To this end, in addition to the previous 
wall-source and point sink setup, we perform various different exercises as described below. 

\vspace*{0.1in}

\noindent{\it{A. Wall-source and Gaussian-smeared sink}:}

\noindent In the first additional setup, we use a wall source along with a Gaussian-smeared 
sink which have been extensively used in heavy hadron calculations \cite{Davies:1994mp,Wurtz:2015mqa}. 
Choosing a suitable Gaussian-width one can achieve a reliable ground state plateauing in 
the correlation functions. We choose the Gaussian width such that the effective energies 
feature a conventional falling-from-above behavior and yet the statistical noise in the correlation 
functions do not wash the signal away. We observe that the best value of Gaussian width across 
all the ensembles we study is $\sim0.2$ fm. Corresponding results obtained on our finest lattice ensemble 
are shown in Fig. \ref{fig:eff_mass_gs}. It is clear from this figure that results from 
the wall-source point-sink ({\it w-p}) correlators and the wall-source Gaussian-smeared sink ({\it w-gs}) correlators 
are quite consistent with each other. The effective mass obtained from the {\it w-gs} correlators for 
the dibaryon operator always stays below than that of the corresponding non-interacting two-baryon 
correlators. We fit these {\it w-gs} correlators with one exponential and the fitted results with 
errorbars are shown by the horizontal bands.
\bef[h]
\centering
\hspace*{-0.06in}\includegraphics[scale=0.45]{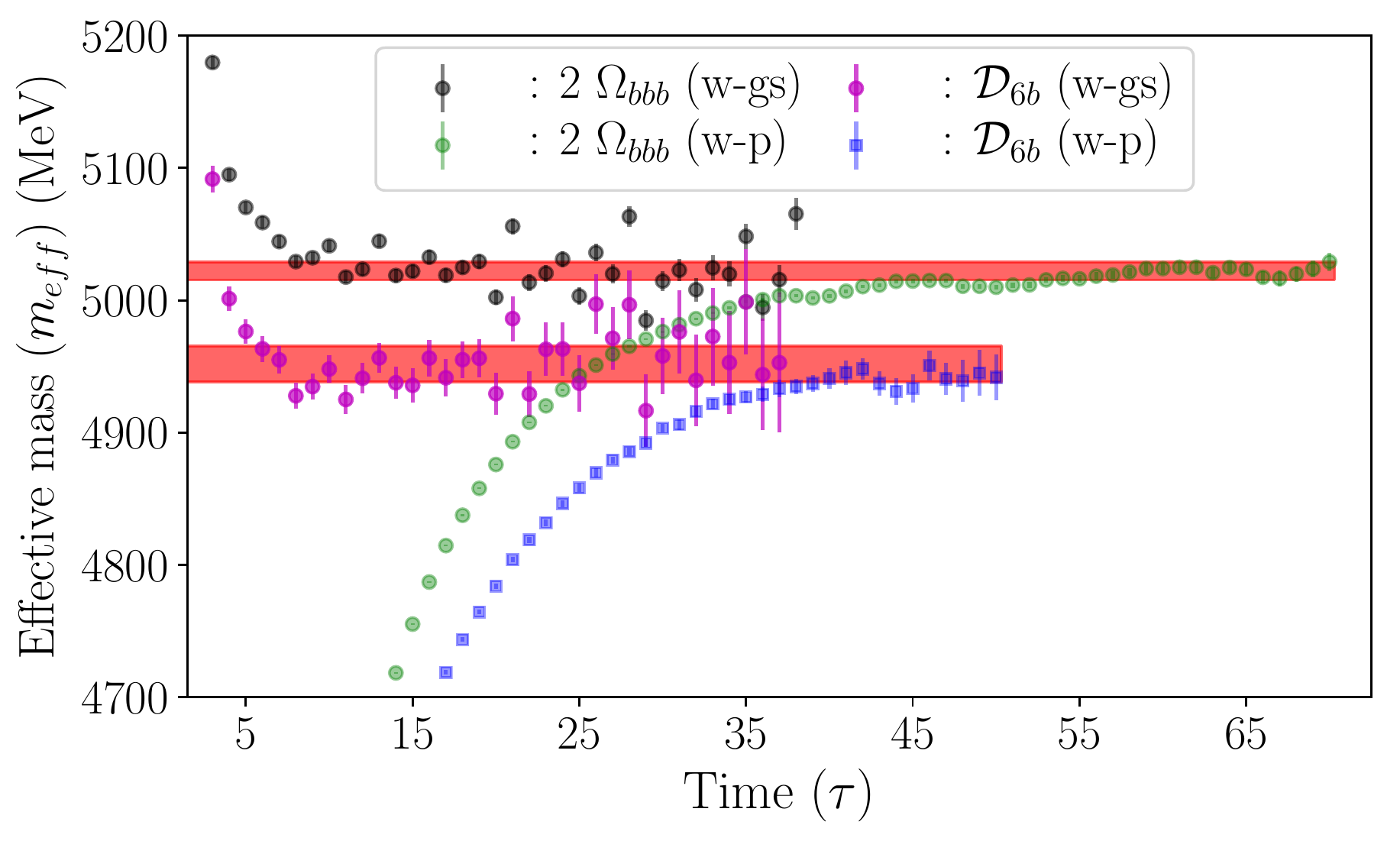}
\caption{\label{fig:eff_mass_gs} Effective mass plots of the  wall-source Gaussian-smeared sink ({\it w-gs} ) 
correlators of the dibaryon and two non-interacting $\Omega_{bbb}$ baryons are shown along with 
the corresponding effective masses from the wall-source point-sink ({\it w-p} ) correlators. Horizontal 
bands show the fitted results for the {\it w-gs} correlators with 1-$\sigma$ error including the systematic 
uncertainties from various possible fitting-windows. The results presented are for
the ensembles with the finest lattice spacing.}
\eef{}

\vspace*{0.1in}

\noindent{\it{B. Wall-source and spherical-box-smeared sink}:}

\noindent We employ a second setup with a spherical-box-smeared sink. This procedure was utilized in Ref. 
\cite{Hudspith:2020tdf} for doubly heavy tetraquark calculations and have been found to be effective in avoiding 
the rising behavior in the effective energies, and in getting an early plateau at the correct ground state energy. 
We have varied the radius ($r$) of the spherical-box and have tuned its value such that the effective energies show 
a falling behavior and yet retains a sufficiently good signal-to-noise ratio. In Fig. \ref{fig:eff_mass_sb}, we 
plot the effective mass obtained on our finest ensemble with $r \sim 0.34$ fm. It is evident that the effective 
mass falls from above and its asymptotic behavior is consistent with that of the wall-point and wall-Gaussian-smeared 
setups that we have discussed. We fit these wall-box ({\it w-b}) correlators and the results with 1-$\sigma$ 
and fitting-window errors are shown by the horizontal bands in Fig. \ref{fig:eff_mass_sb}.

\bef[h]
\centering
\hspace*{-0.06in}\includegraphics[scale=0.45]{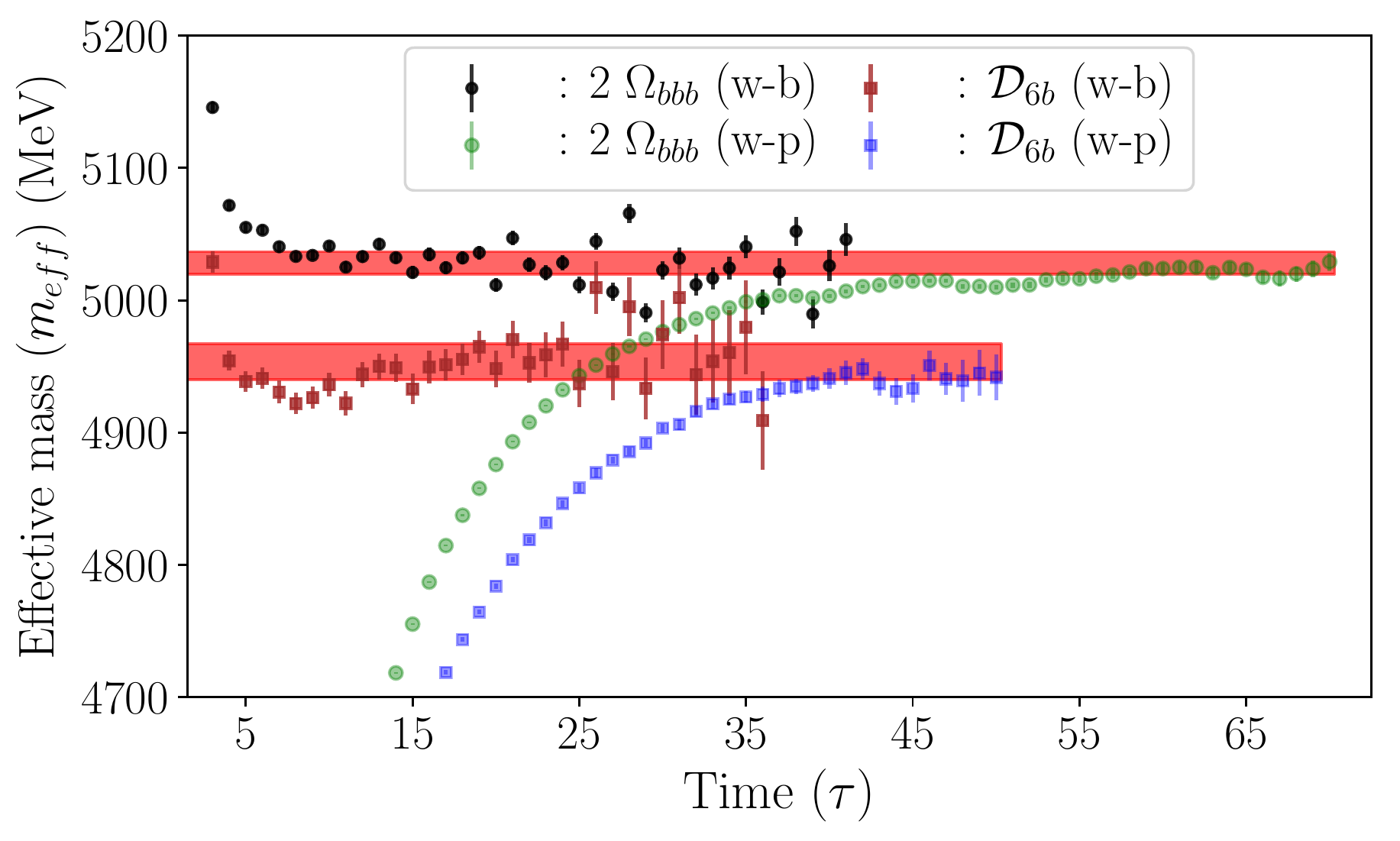}
\caption{\label{fig:eff_mass_sb} Effective mass plots of the wall-source spherical-box-sink  ({\it w-b})
correlators for the dibaryon and two non-interacting $\Omega_{bbb}$ baryons are shown along with 
the corresponding effective masses from the wall-source point-sink (w-p) correlators. The bands show 
the fitted results for the {\it w-b} correlators with 1-$\sigma$ error including the systematic uncertainties 
from various possible fitting-windows. The results presented are for the ensembles with the finest lattice spacing.}
\eef{}

\vspace*{0.15in}

\noindent{\it{C. Effective mass using Prony's method}:}

\noindent Further to the above consistency checks, we have performed another complementary analysis known as the ``Prony's-method'' 
\cite{Prony1795,Fleming:2004hs,Kunis2015AMG} to find the reliability of the ground state plateau and the extent of excited state contaminations in the ground state energy estimation. 
Although this method was found to be unstable with smaller statistics, it was shown to be quite 
effective to get a reliable ground state effective mass with the high statistics correlation 
functions \cite{Beane:2009kya}. It was also found to produce the energy levels favorably to 
that obtained through the variational approach \cite{Lin:2007iq}. In Fig. \ref{fig:prony} we 
show the effective mass obtained for the wall-point correlators using this method with two exponentials where the solutions 
are numerically stable (solving Eq. (16) and (17) of Ref. \cite{Beane:2009kya} numerically). We also find that the solutions are often unstable with large errors for arbitrary choices of $n$ and $q_1$ values of 
Eq. (16) and (17) of Ref. \cite{Beane:2009kya}.  Using suitable choices of $n$ and $q_1$ we find stable solutions and show that in Fig. \ref{fig:prony}. The effective masses are 
shifted towards the right as per the choice of $n$ and $q_1$. It is clear 
from this plot that the effective masses obtained using one exponential from the wall-point 
correlators are consistent with that obtained using Prony's method using two exponentials. 
This provides another consistency check of the findings using wall-point correlators.

\bef[h]
\centering
\hspace*{-0.06in}\includegraphics[scale=0.45]{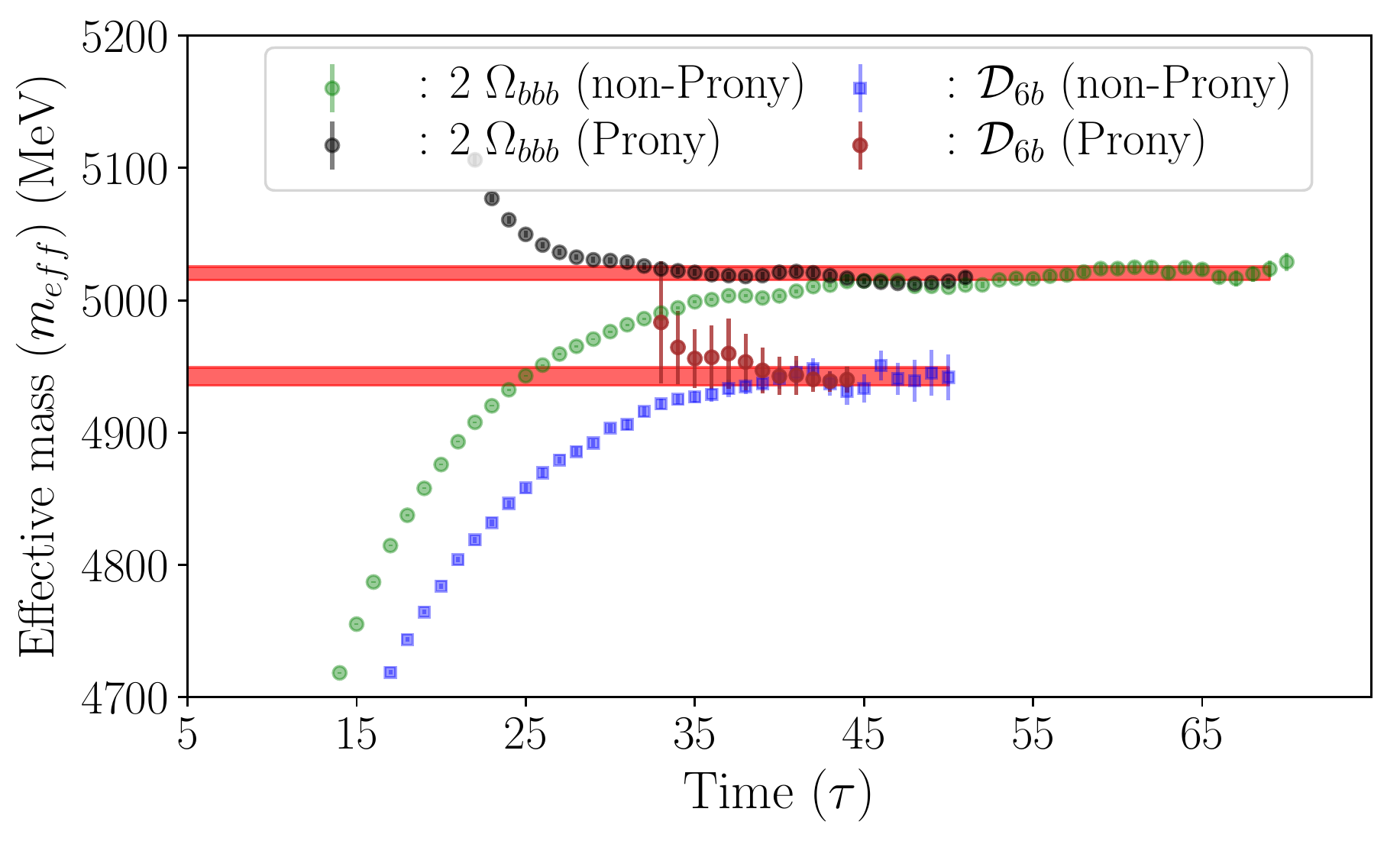}
\caption{\label{fig:prony} Effective mass obtained from using Prony's method \cite{Prony1795,Fleming:2004hs,Kunis2015AMG}. The band shows the fitted results with its error determined using the wall-source point-sink data.}
\eef{}

\vspace*{0.1in}

\noindent{\it D. Dibaryon operators with displaced baryons at sink}:
  
\noindent As an additional exercise, we investigate the effect of using displaced baryons ($B$) 
for two-baryon operators $O(r=0; \overline x) = B(\overline x)B(\overline x)$ at the sink, and also 
compare the observations with that obtained with similar setup for the well established deeply 
bound system of doubly heavy tetraquarks. To this end, we displace the two baryons in the two-baryon 
operator such that $O(r;(\overline x_1 + \overline x_2)/2) = B(\overline x_1)B(\overline x_2)$, 
where the displacement $r=|\overline x_1 - \overline x_2|$ (here $r$ is symmetrized with respect 
to all three spatial directions). We observe that the ground state energy estimate from such an 
operator is consistent with that of the local two-baryon operators until below $r\sim0.25$ fm. 
Note that this also coincides with the chosen width of the Gaussian smearing and closer to 
the boundary of the spherical-box-smearing that yields a conventional behavior of falling 
effective energies. In Fig. \ref{fig:shift}, we present these results obtained on the finest ensemble with different values of 
$r$.

\bef[h]
\centering
\hspace*{-0.06in}\includegraphics[scale=0.45]{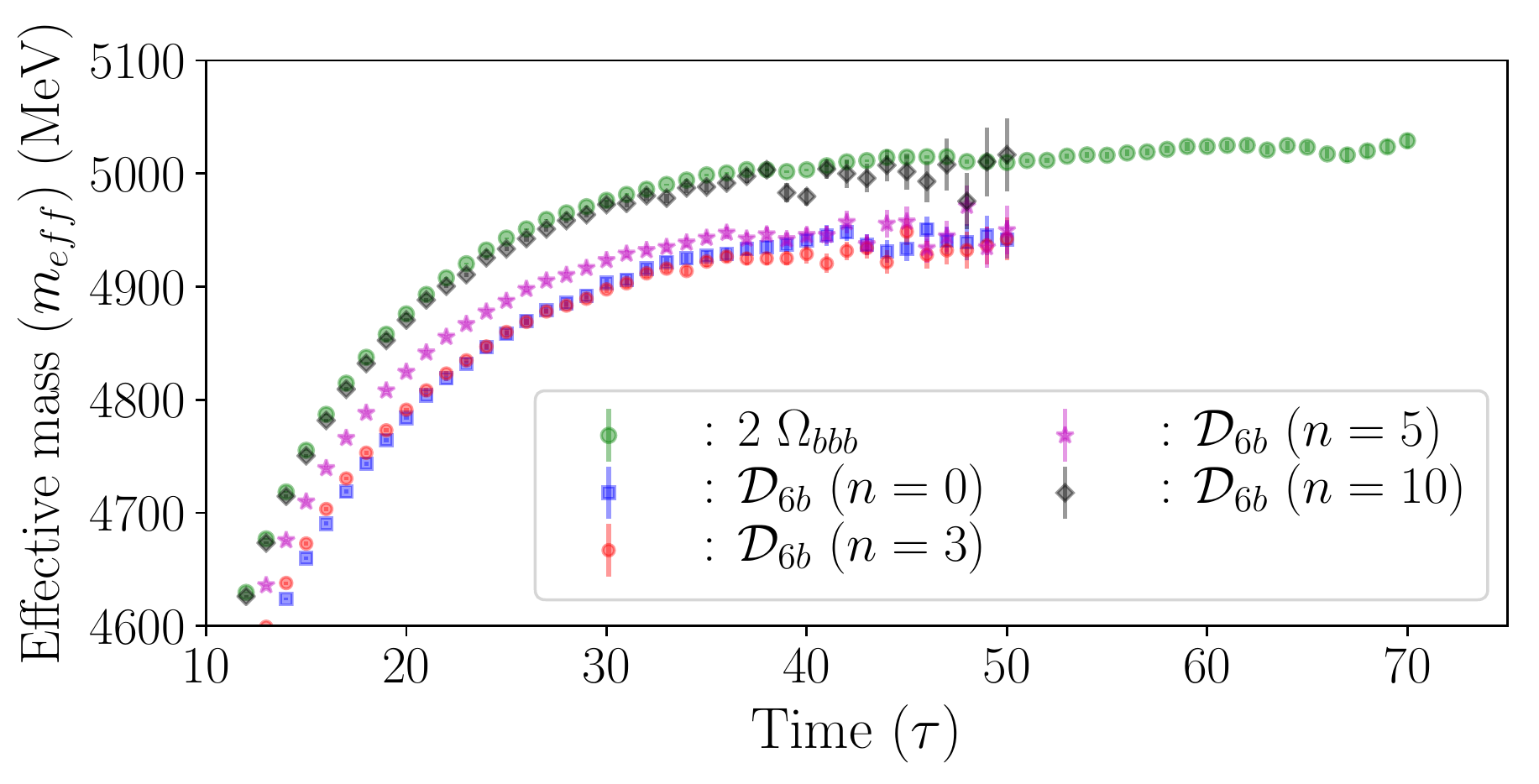}
\caption{\label{fig:shift} Effective masses corresponding to $2\Omega_{bbb}$, $D_{6b}$ with 
and without displacement at the sink points (see text for more explanation). The two baryon 
operators are built with displaced baryons at the sink with $r = na$ referring to the displacement 
between the baryons. The results presented are for the ensembles with the finest lattice spacing.}
\eef{}

To understand the significance of this result, we then carried out the same exercise for the case of 
well-studied doubly-bottomed tetraquarks ($\bar{b}\bar{b} u d$) with the same asymmetric wall-source and 
point-sink setup. Note that for this case, the energy level corresponding to the four-quark operator 
of two-meson type lies below than that of the threshold level obtained from the two non-interacting 
two mesons. As in the case of dibaryon, we vary the two sink points ($\overline x_1 $ and $\overline x_2$) 
of two-meson operators ($\bar{b}(\overline x_1) u(\overline x_1)\bar{b}(\overline x_2) d(\overline x_2)$). 
Interestingly, we find strikingly similar results as above, where up to a certain displacement ($r$), 
the effective mass of two-meson operators with one ($\overline x_1 = \overline x_2$) and two sink 
points ($\overline x_1  \ne \overline x_2$) are found to be consistent with each other. 
The $\bar{b}\bar{b} u d$ systems have been studied recently with variational methods by 
multiple lattice-QCD groups with asymmetric \cite{Francis:2016hui,Junnarkar:2018twb}, box-smeared 
\cite{Hudspith:2020tdf} as well as smeared point-to-all correlators \cite{PhysRevD.100.014503}, and 
results on the bindings obtained by those different lattice calculations consistently found a deeply 
bound state. Since the response to the binding with respect to the displacement of two sink points 
are strikingly similar both for the dibaryon studied here and for 
$\bar{b}(\overline x_1) u(\overline x_1)\bar{b}(\overline x_2) d(\overline x_2)$
tetraquarks, and  $\bar{b}\bar{b} u d$ was found to be deeply bound by multiple studies, we believe our result is robust up to the statistical and systematic uncertainties that we have determined. This finding on the existence of a deeply bound heavy dibaryon calls for further lattice calculations on heavy dibaryons particularly using  multi-operator variational approaches to quantify the systematics related to the hard-to-quantify excited state effects more precisely, which have been found to affect the results obtained for light dibaryons that employed asymmetric correlators \cite{Amarasinghe:2021lqa,Green:2021qol}.

In summary, various procedures followed above led us to conclude that the results obtained using 
asymmetric wall-point setup is robust and the effect of the excited state is minimum as long as 
the fitting window corresponds to the real plateau from the ground state energy. We find that the 
ground state plateau saturates $\gtrsim2$ fm for both the baryon and dibaryon correlators across 
all the ensembles. We also observe that the dibaryon correlator overlaps with the ground state maximally 
when the smearing size is about 0.2 fm. This is also in line with the results from two-baryons operators 
with displaced baryons at the sink. This observation on the smearing width ($\sim 0.2$ fm) perhaps is 
indicating that the observed dibaryon could be a compact state. To account for the effect of  
contaminations from the excited state we include conservative errors  
determined based on the difference in energy estimates obtained from different calculations, as discussed above.

\vspace*{0.1in}

\noindent{3. \it{Continuum extrapolation}:}

\noindent We employ a set of lattice QCD ensembles in which gauge fields are Symanzik-improved at $\mathcal{O}(\alpha_sa^2)$ 
and include the effect of $u$, $d$, $s$ and $c$ quark vacuum polarization generated with the highly improved 
staggered quark action \cite{Bazavov:2012xda}. Quark propagators are generated with NRQCD action with improvement 
coefficients up to $\mathcal{O}(\alpha_sv^4)$. The lattice spacing dependence of the energy differences (in Table I)
could be nontrivial. Similar to the approach made in Ref.~\cite{Green:2021qol}, we account for this by parameterizing $kcot\delta_0$, 
that enter the scattering analysis in Eq. 5, with different forms and perform fits with different sets of 
energy levels determined from the simulation. Choosing the linear parameterization 
$k~cot\delta_0 = -1/a_0^{[0]} - a/a_0^{[1]}$ that best describes the entire data, we find the total uncertainties arising 
from statistics, fitting window and continuum extrapolation to be $\sim$18\% of the binding energy from the continuum extrapolation.
We find that choosing other forms of continuum extrapolation for the scattering length $-1/a_0$ leads to a change 
of at most 8 MeV in the binding energy, which we quantify as the uncertainty arising from the discretization error.

\vspace*{0.1in}

\noindent{4. \it Scale setting}:

\noindent Scale settings through $r_1$ parameter \cite{Bazavov:2012xda} and Wilson-flow were 
found to be consistent \cite{Bazavov:2012xda} for these lattice ensembles. Systematics with the scale settings 
further gets reduced in the estimation of energy differences (Eq. 4), and as in Ref. 
\cite{Mathur:2018epb,Junnarkar:2019equ} we find it to be maximum of about 3 MeV.

\vspace*{0.1in}

\noindent{5. \it Quark mass tuning}:

\noindent We tune the bottom quark mass employing the Fermilab method of heavy quarks \cite{ElKhadra:1996mp}. Here, 
we equate the  lattice extracted spin average $\overline{1S}$ bottomonia {\it kinetic mass}, 
${1\over 4} [3 M_{\Upsilon} + M_{\eta_b}]_{kin}$, with its physical value. We perform this tuning corresponding 
to the central value of the chosen scale and also at its error values. We calculate $E_{\mathcal{D}_{6b}}$ 
for each of the tuned masses and include the variation as the estimation of error due to quark mass tuning. 
We find it to be less than 2 MeV.

With the above mentioned lattice setup we find the hyperfine splitting in $1S$ bottomonia, a benchmark observable 
for the evaluation of the goodness of lattice calculations with bottom quarks, is quite consistent with its experimental value, as demonstrated in Figure \ref{fig:hfs}. The continuum value (green star) is obtained taking the average of estimates from all ensembles and the error (green band) is estimated as a weighted average with respect to the lattice spacings. Continuum extrapolation with the linear as well as and quadratic forms in lattice spacing are also 
shown by the orange and blue stars respectively with the same color bands for their 1-$\sigma$ errors. Together with 
possible other systematics, that we are discussing here, we estimate its value to be 62.6(3)(5) MeV.

\bef[h]
\centering
%
\hspace*{-0.1in}\includegraphics[scale=0.5]{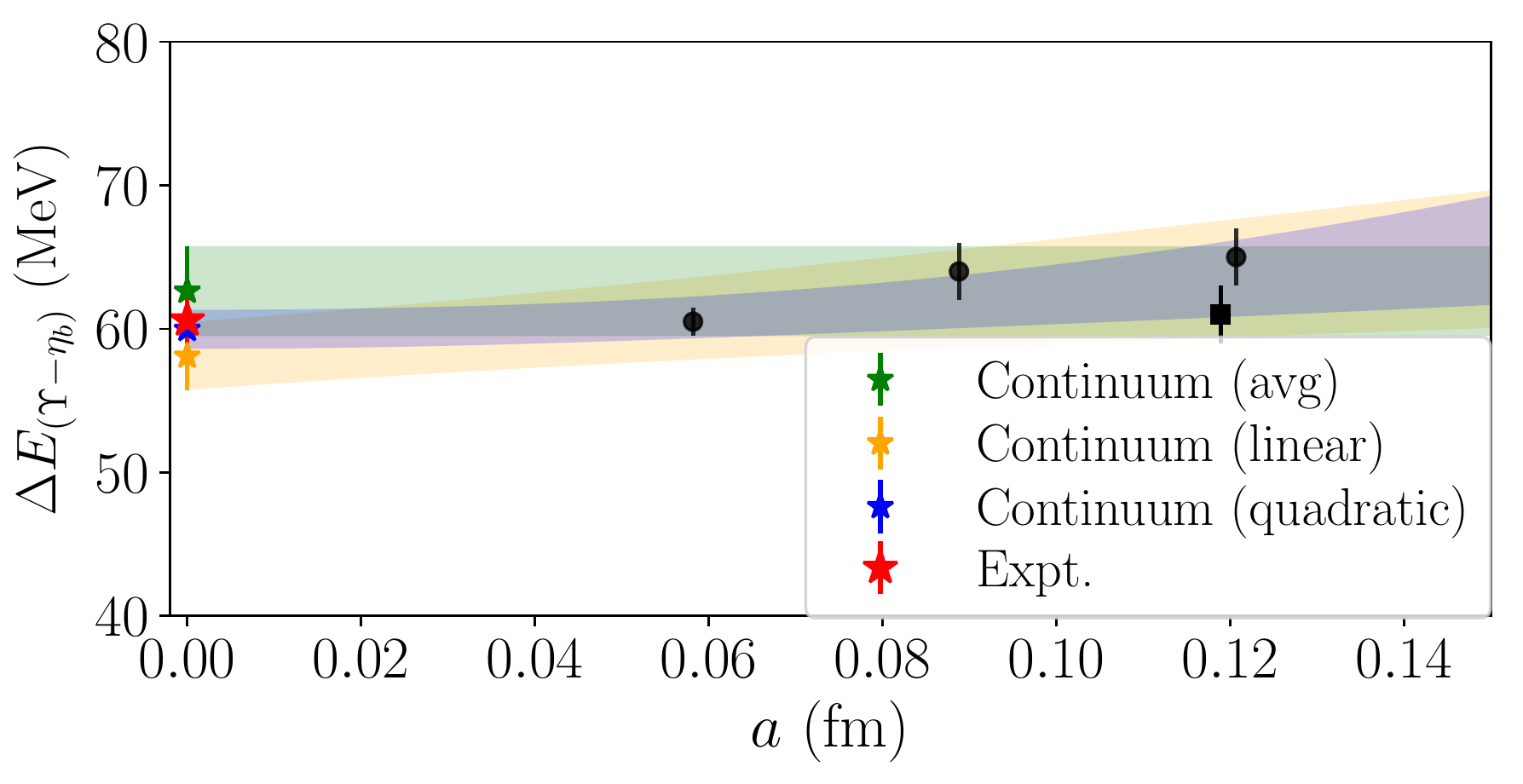}
\caption{\label{fig:hfs} Hyperfine-splitting of the $1S$ bottomonia.}
\eef{}

\vspace*{0.1in}

\noindent{6. \it Electromagnetism}:

\noindent The dibaryon investigated here has two units of electric charge which can affect its binding. To gauge that, we 
perform the following analysis as in Ref. \cite{Lyu:2021qsh}. First, we model the strong interactions 
between two interacting $\Omega_{bbb}^{-}$ baryons with a quantum mechanical multi-Gaussian attractive potential $V_s$ \cite{Lyu:2021qsh}, constrained  to match the binding energy $-81(_{-16}^{+14})$ MeV that we find in this work. 
Next, we assume the form of the Coulomb potential ($V_e$) of $\Omega_{bbb}^{-}$ to be similar to that of $\Omega_{ccc}^{++}$, 
except the total electric charge is $-2$. We present a comparison of the strengths of these potentials as a function of the radial distance in Figure 4 of main text, with the root-mean-square ({\it rms}) charge radius $r_d$ chosen as the {\it rms} radius of the ground state of $V_s$. Next, we solve the energy eigenvalue problem with
the effective potential ($V_{eff}=V_s+V_e$) and determine the
scattering length $a^{e+s}_0$ and effective range $r^{e+s}$, following the procedure discussed in Ref. \cite{Lyu:2021qsh}. 
The radial probability densities of the ground state wave-functions (dashed-dotted curves) corresponding to $V_s$ and $V_{eff}$ are shown in Figure 4 of the main text. It is evident that the Coulomb repulsion serves only as a perturbation and hence does not change the binding energy of $\mathcal{D}_{6b}$ 
in any significant way. We also vary $r_d$ and find that the effect of Coulomb repulsion is largely perturbative and binding may reduce at most by 10 MeV even when $r_d$ is chosen to be unphysically low as 0.01 fm. We present $1/a^{e+s}_0$ for $V_{eff}$ as a function of the Coulomb interaction strength $\alpha^{e}$ in Figure \ref{fig:coul2}. Note that $1/a^{e+s}_0$ remains to be very much {\it positive} even at $\alpha^{e} = \alpha^{e}_{phys}$, confirming that $\mathcal{D}_{6b}$ remains to be a deeply bound state even in the presence of Coulomb repulsion, with a total binding energy of about $ -75 $ MeV.
\bef[h]
\centering
\includegraphics[scale=0.18]{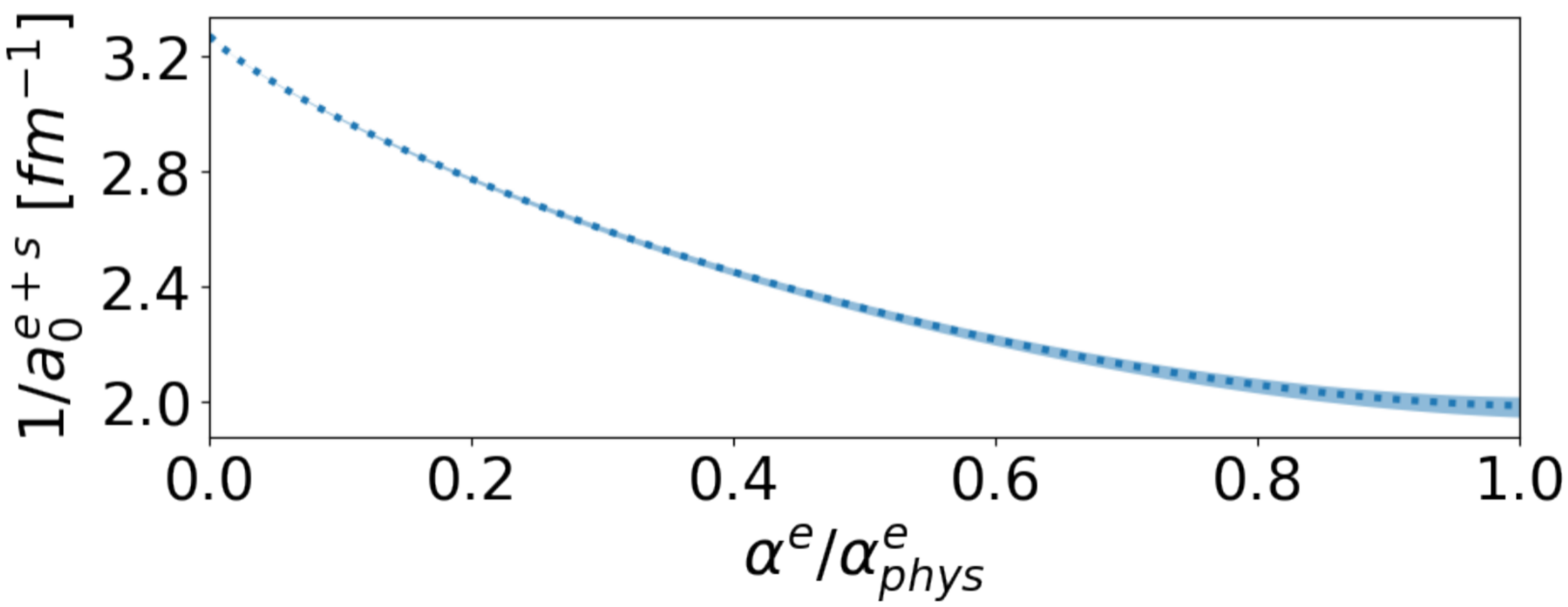}
\caption{\label{fig:coul2}The inverse of the scattering length $1/{a^{e+s}_0}$
as a function of $\alpha^{e}/\alpha^{e}_{phys}$. Its large {\it positive} value throughout indicates that $\mathcal{D}_{6b}$ is a deeply bound state even in the presence of Coulomb repulsion.}
\eef{}

For heavy baryons, the possible systematics due to other electromagnetic corrections was found to be 3 MeV \cite{Borsanyi:2014jba}. Keeping that in mind as the 
source of other electromagnetic effects beside the Coulomb repulsion, we take a conservative estimate of 
8 MeV corrections for the binding energy (by adding the average of Coulomb repulsion with the above mentioned 3 MeV in quadrature).

No chiral extrapolation is necessary for $\mathcal{D}_{6b}$. For heavier dibaryons the unphysical sea 
quark mass effects are expected to be within a percent level ~\cite{McNeile:2012qf, Dowdall:2012ab, Chakraborty:2014aca}, 
and particularly for $\mathcal{D}_{6b}$, it would be negligibly small. In Table \ref{error-table} we summarize the error-budget estimate where above mentioned systematics are added in quadrature.

\bet[h]
\centering
\begin{tabular}{l|c }
\hline
\hline
$Source$ & Error (MeV)\\
\hline
Statistical + Fit-window + & \multirow{2}{*}{\large$\left(^{+16}_{-14}\right)$}\\
Continuum extrapolation\\\hline
Excited states & 8\\
\hline
Discretization & 8\\
  Scale setting& 3   \\
  $m_b$ tuning & 2  \\
  Electromagnetism & 8\\
  \hline
  Total systematics &  12 \\
  \hline
\hline
\end{tabular}
\caption{\label{error-table}Error budget in the calculation of the binding energy $\Delta E_{\mathcal{D}_{6b}}$. The total systematics quoted above includes those from the discretization, scale setting, bottom quark mass 
tuning and electromagnetic effects.}
\end{table}


\end{document}